\documentclass[11 pt]{article}
\usepackage{amsmath,amssymb,graphicx}
\usepackage[colorlinks=true]{hyperref}
\renewcommand{\u}{\underline}
\newcommand{\ket}[1]{\left | #1 \right\rangle}
\newcommand{\bra}[1]{\left \langle #1 \right |}
\newcommand{\braket}[2]{\left\langle #1|#2\right\rangle}

\newcommand{\dg}{\dagger}
\renewcommand{\u}{\underline}
\def\identity{\leavevmode\hbox{\small1\kern-3.8pt\normalsize1}}
\def\er{{\cal E}}

\begin{document}

\begin{center}
{\Large{\bf{How to Counteract Systematic Errors in Quantum State Transfer}}}
\bigskip\bigskip 

{{{\bf{Chiara Marletto}}$^{1}$, {\bf{Alastair Kay}}$^{2,3}$, and {\bf Artur Ekert}$^{1,2}$}}

\medskip 
\small {$^{1}$ Mathematical Institute, 24-29 St Giles', Oxford OX1 3LB, UK }\\
\smallskip
\small{$^2$Centre for Quantum Technologies, National University of Singapore, 3 Science Drive 2, Singapore 117543}\\
\small{$^3$Keble College, Parks Road, Oxford, OX1 3PG, UK}\\
\bigskip
{\bf Abstract}\\
\end{center}
In the absence of errors, the dynamics of a spin chain, with a suitably engineered local Hamiltonian, allow the perfect, coherent transfer of a quantum state over large distances. Here, we propose encoding and decoding procedures to recover perfectly from low rates of systematic errors. The encoding and decoding regions, located at opposite ends of the chain, are small compared to the length of the chain, growing linearly with the size of the error. We also describe how these errors can be identified, again by only acting on the encoding and decoding regions.

%
%\title{How to Counteract Systematic Errors in Quantum State Transfer}
%%
%
%\author{Chiara Marletto}
%\affiliation{Mathematical Institute, University of Oxford, 24-29 St Giles', OX1 3LB, UK}
%\author{Alastair Kay}
%\affiliation{Centre for Quantum Technologies, National University of Singapore, 3 Science Drive 2, Singapore 117543}
%\affiliation{Keble College, Parks Road, Oxford, OX1 3PG, UK}
%\author{Artur Ekert}
%\affiliation{Mathematical Institute, University of Oxford, 24-29 St Giles', OX1 3LB, UK}
%\affiliation{Centre for Quantum Technologies, National University of Singapore, 3 Science Drive 2, Singapore 117543}
%
%\begin{abstract}

%%\end{abstract}
%\maketitle

\section{Introduction}
Quantum state transfer, i.e., the coherent transfer of an unknown quantum state between two distant locations, plays a fundamental role in quantum information processing, facilitating the production of entanglement between separated parties, and generating the long-range two-qubit interactions required in a quantum computer. If the available inter-qubit interactions only act locally, one of the most effective ways to achieve this state transfer is to use a spin chain as a quantum wire \cite{BOS, CHDADO, CHDAEK} (see \cite{KAY} for a review). In this scheme, the dynamics of a suitably designed Hamiltonian perfectly transfer a quantum state between the two ends of an arbitrarily long chain. Its main strength is that external control is not required during the transfer while confining the encoding and decoding areas to the boundary regions of the chain, so that it is far superior to the na\"ive solution of applying a sequence of SWAP gates. Since the protocol is designed to operate in the same solid-state technology that the rest of the quantum device is fabricated in, interfacing with other technologies, such as using photons as `flying qubits' \cite{CIZOKI}, is unnecessary. 

One of the promising features of a spin chain is that, in only requiring interactions with the ends of a chain, the majority of the spins can be shielded from the effects of noise. Nevertheless, errors will inevitably be generated in a practical implementation by manufacturing imprecisions and through coupling with the environment. The main effect of these errors is to compromise the transfer by introducing unwanted excitations and destroying the coherence. In order to reduce this effect, one can consider encoding in an error correcting code. However, applying standard error-correcting codes to state transfer is very inefficient; any error occurring during the transfer, even on a single qubit, is spread over the whole chain by the action of the Hamiltonian. Hence, one expects that a generic error-correcting code would require a large number of encoding qubits per logical qubit, conflicting with the central tenet of the state transfer protocol. It is therefore necessary to design {\em ad hoc} error correcting codes. In addition to the importance of error correction on the spin chain itself, such studies comprise the essential initial studies of equivalent questions in systems where Hamiltonian dynamics are used in an information processing capacity, such as in a computational architecture \cite{KAPE}.

 To date, studies of error correction in state transfer have adopted a cause-based approach.  Manufacturing errors have been modelled as random static or dynamical defects, \cite{OSB, BUEIOS}, or studied via simulations \cite{RON}, and can be circumvented (at the cost of a significantly increased transfer time) by encoding across multiple parallel chains \cite{BOBUR, BOSBUR}. The effects of timing errors can be minimised by optimally tuning the couplings between the chain qubits \cite{KAY2}. The studies of environmental induced noise models are, to date, extremely limited, being based around the very crude attempts such as in \cite{KAY, KAY3}, or with very specific models with little physical motivation, \cite{WIE}. Nevertheless, there are specific occasions when perfect state transfer can be recovered, albeit with a doubled transfer time \cite{BUBO}. No general procedure to protect against noise is yet known; indeed, aside from some of the arguments of \cite{KAY}, all these prior results have made use of a deliberately induced subspace structure in the Hamiltonian which is assumed to be preserved, even in the presence of these errors\footnote{This is usually, but not exclusively, the spin preserving property that we describe in Sec.\ \ref{sec:basics}.}. One promising avenue, recently proposed in \cite{CIZO}, advocates the use of topologically protected subspaces in chiral spin liquids in order to protect against decoherence during the state transfer.

This cause-based classification has so far concealed an interesting category of errors, {\em systematic errors}, by which we mean errors that in each transfer experiment are drawn from a fixed set of possible errors (whose description includes position, time, and type of error). This key feature enables the detection, once and for all, of the set of possible errors, potentially allowing us to protect against them. In this paper, we develop a strategy to recover perfectly from low rates of systematic error on a spin chain, encoding the information in a protected set of states and decoding it accordingly. The number of qubits used for the encoding and decoding scales linearly with the sizes of the errors in the set. We start, in Sec.\ \ref{sec:basics}, by introducing the requisite information about spin chains, particularly concentrating on the mapping to a system of free fermions, the mathematical structure which we rely on throughout the paper. The encoding and decoding strategies are presented in Section \ref{sec:main} under the assumption that the errors are known, before we describe an example of how the errors can be identified in Section \ref{sec:determine}, while also indicating that significant efficiency savings can be made over and above the sufficient conditions previously derived.

\subsection{Spin Chains and Fermions} \label{sec:basics}

In an ideal state transfer protocol, we are provided with an unknown quantum state
$$
\ket{\chi}\doteq (\alpha\ket{0}+\beta\ket{1}),
$$
and tasked with transmitting this state perfectly to a distant recipient, using only a fixed Hamiltonian with local couplings. The most commonly studied system is that of a spin chain, initially because this is the optimal geometry for transferring a state as far as possible. However, the many advantages of this choice include the ability to prove necessary and sufficient conditions for achieving perfect state transfer \cite{KAY}, and the ability to perform the Jordan-Wigner transformation on the system \cite{ALBA}. We will also take advantage of these benefits.

In the simplest scenario, this state transfer is achieved by placing the state onto the first qubit of an $N$ qubit chain, and initialising the rest of the system in a standard state such as $\ket{0}^{\otimes N-1}$. Thus, the overall starting state is
$$
\ket{\psi_{I}}=\ket{\chi}\otimes\ket{0}^{\otimes N-1}
$$
By suitably selecting the properties of the Hamiltonian $H$, we evolve for some time, $t_f$, known as the transfer time, and the corresponding unitary $U(t_f)=e^{-iHt_f}$ creates the final output
$$
\ket{\psi_{F}}\doteq 
U(t_f)\ket{\psi_{I}}=\ket{0}^{\otimes N-1}\ket{\chi},
$$
i.e.\ the state arrives on qubit $N$ of the chain. We consider Hamiltonians of the class
$$
H=\frac{1}{2} \sum_{i=1}^{N-1}J_i(X_iX_{i+1}+Y_iY_{i+1})-\sum_{i=1}^{N}B_iZ_i
$$
where $X_i, Y_i$ and $Z_i$ are the Pauli operators acting on the $i$-th qubit. This Hamiltonian has a very convenient subspace structure, described by
$$
\left[H,\sum_{i=1}^NZ_i\right]=0,
$$
meaning that the total number of ones (excitations) relative to the state
$$
\ket{\bf 0}=\ket{0}^{\otimes N}
$$
is preserved. In particular, the one-excitation subspace, spanned by the states
$$
\ket{\u n}=\ket{0}^{\otimes(n-1)}\ket{1}\ket{0}^{\otimes(N-n)},
$$
is preserved. This allows us to describe the dynamics of this subspace by a Hamiltonian $H_1$, where $(H_1)_{n,m}\doteq\bra{\u n}H\ket{\u m}$. Perfect transfer is achieved by an appropriate choice of coefficients $J_i$ and $B_i$, the necessary and sufficient conditions for which are described in \cite{KAY}, which result in the following properties:
\begin{eqnarray*}
U(t_f)\ket{\bf 0}&=&\ket{\bf 0} \\
U(t_f)\ket{\u n}&=&\exp(i\phi)\ket{\u {N-n+1}}\qquad\forall n=1\ldots N
\end{eqnarray*}
for some real time $t_f$ and phase $\phi$. There are many such choices of coupling that achieve perfect state transfer. While the original \cite{ CHDADO, CHDAEK} has been shown to be optimal against a variety of criteria \cite{KAY2}, we remain agnostic to the specific choice, and do not require more than the basic properties outlined here. The phase $\phi$ shall be neglected in the following, as it can be corrected by a local rotation.

As we will be concerned by the action of errors which, generically, may introduce additional excitations during the state transfer, we shall switch to the fermionic picture, which lends itself to the description of variable excitation number. In this picture, the Hamiltonian reads:
$$
H= \sum_{n=1}^{N-1}J_{n}(a_{n}^{\dagger}a_{n+1}+a_{n+1}^{\dagger}a_{n})+ 2 
\sum_{n=1}^{N}B_{n}a_{n}^{\dagger}a_{n}
$$
where
$$
\{a_n^{\dagger},a_m\}=\delta_{n,m}\qquad\{a_n, a_m\}=0.
$$
This hopping Hamiltonian describes $N$ non-interacting spinless fermions \cite{KAY}. The Jordan-Wigner transformation $\displaystyle{a_n^{\dagger}=\frac{1}
{2}\prod_{m=1}^{n-1}Z_m\;\big (X_n-iY_n\big)\;}$, $Z_n=\identity-2a_n^{\dagger}a_n$ provides the relation to the spin 
picture. Setting $\hat {\cal O}(t)\doteq U(t){\cal O}U(t)^{\dg}$, for any operator $\cal O$,
the bilinear structure of $H$ implies that
$$
\hat a_n(t)=\sum_{m=1}^{N}\beta_{n,m}^{*}(t)a_m,
$$
where $\beta_{n,m}(t)\doteq \bra{n}\exp(-iH_1 t)\ket{m}$ \cite{JOMI} (i.e., the dynamics are entirely determined by the single excitation subspace, described by $H_1$). Hence,
two fermions are perfectly transferred independently of one another, modulo an ordering phase: 
$$
U(t_f)a_na_mU^\dagger(t_f)=a_{N-n+1}a_{N-m+1}.
$$ 

\subsection{Initial State} \label{sec:init}

We specified that for the basic state transfer protocol, the initial state of every qubit except the first one should be $\ket{0}$. However, a relatively minor adaptation of the protocol makes this requirement unnecessary \cite{KAY3,PAT2}. Consider, instead, fixing the first two qubits to convey the information of the unknown state via an encoding
$$
(\alpha a_1^\dagger+\beta a_2^\dagger)\ket{00},
$$
and the state of the rest of the chain can be completely arbitrary. The state of the rest of the chain is described by an arbitrary state
$$
\left(\sum_{x\in\{0,1\}^{N-2}}\gamma_xa_x^\dagger\right)\ket{0}^{\otimes N-2},
$$
where $a_x^\dagger\doteq \prod_{i=3}^{N}(a_i^{\dagger})^{x_i}$ uses the bits of the binary string $x$ to specify which of the $N-2$ qubits have creation operators acting on them. Hence,
$$
\ket{\psi_I}=(\alpha a_1^\dagger+\beta a_2^\dagger)\ket{00}\left(\sum_{x\in\{0,1\}^{N-2}}\gamma_xa_x^\dagger\right)\ket{0}^{\otimes N-2},
$$
but by the properties of perfect state transfer and the non-interacting fermion picture, we immediately have that the output state is
$$
\ket{\psi_F}=\left(\sum_{x\in\{0,1\}^{N-2}}\gamma_{x^R}(-1)^{w_x+\binom{w_x}{2}}a_x^\dagger\right)\ket{0}^{\otimes N-2}(\alpha a_N^\dagger+\beta a_{N-1}^\dagger)\ket{00},
$$
where $x^R$ is the bit string $x$ reversed and $w_x$ is the Hamming weight of $x$, i.e.\ the number of creation operators that act in $a_x^\dagger$. Clearly, this state is separable about the partition $(1\ldots N-2)$ and $(N-1,N)$, so we can recover the transferred state from the other end of the chain. By linearity, any mixture of these initial states can also be tolerated.

\section{Modelling systematic errors} \label{sec:moder}

 We represent a systematic error by a set of possible errors $\{\mathcal{E}_{\ell }\}$ acting at times $\{t_{\ell}\}$. Since the $\{a_{i}^\dagger\}$ form a complete basis, each of these $\mathcal{E}_{\ell}$ can be expanded as
\begin{equation}
{\cal E} \doteq \sum_{i=1}^{s}\gamma_i \prod_{j=1}^{m_i}a_{k^{(i)}_j}^{\dagger}\prod_{j=m_i+1}^{n_i}a_{k^{(i)}_j}\; ,\;\label{error}
\end{equation}
where $k_j^{(i)}$ are the $n_i$ different positions that the creation/annihilation operators act on. The $\{\gamma_i\}$ are arbitrary coefficients. The overall error $\mathcal{E}$ is taken to be trace preserving. For simplicity of exposition, we shall work with just one error $\mathcal{E}$ acting at time $t$, the effect of multiple errors being easily incorporated. 
The error is assumed to act instantaneously, or, at least, for a time much smaller than $t_f$, so that otherwise the time-evolution is governed by $U$ \footnote{We comment on the relaxation of this assumption in the conclusions.}.
Hence the state at time $t_f$ is $\ket{\psi_{F}}= \hat \er (t_f-t) U(t_f)\ket{\psi_{I}} \;,$ with 
\begin{eqnarray}
\hat \er(t_f-t)=\sum_{i=1}^{s}\gamma_i\prod_{j=1}^{m_i}\hat a_{k^{(i)}_j}^{\dagger}(t_f-t)\prod_{j=m_i+1}^{n_i}\hat a_{k^{(i)}_j}(t_f-t)\;.\label{EVER}
\end{eqnarray}
The fermionic operators $\hat a_i(t_f-t)$'s represent the (possibly non-orthogonal) fermionic modes affected by the error.

Even without knowing the specific form of the error, the encoding presented in Sec.\ \ref{sec:init} already tolerates a vast rage of errors; those for which all the errors $\mathcal{E}$ acting at time $t$ satisfying $\mathcal{E}(t_f-t)=\sum_{x^R}a^\dagger_{x^R}$ (or any mixture thereof), if $\ket{\bf 0}$ is used as initial state, since this is entirely equivalent to the case of the different initial state. Our task is now to describe how to deal with errors that do not satisfy this property, under the assumption that we know what these errors are.

\section{How to counteract systematic errors} \label{sec:main}

\begin{figure}[!t]
 \centering
   {\includegraphics[scale=0.4]{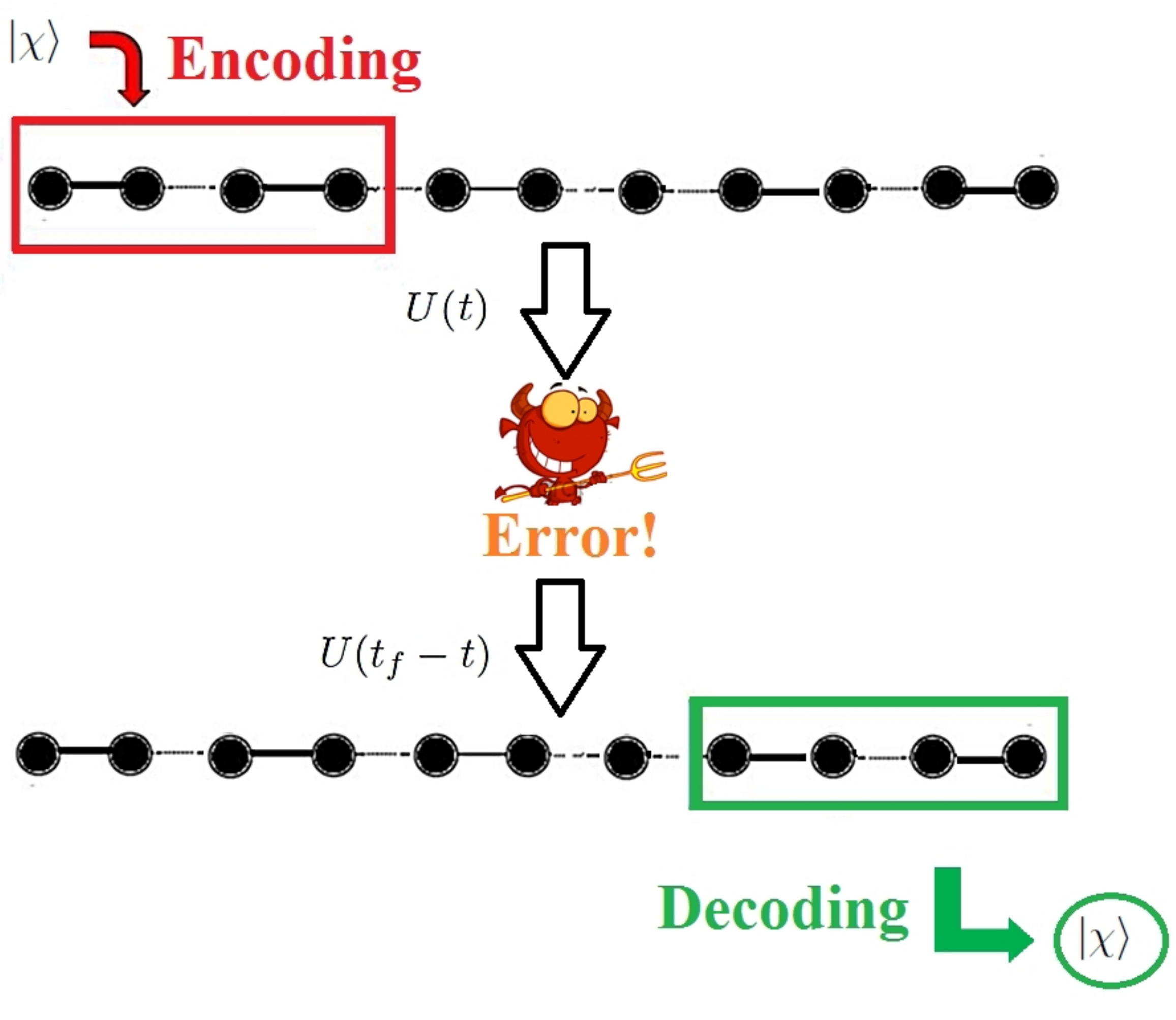}}
\hspace{2mm}
\caption{General schematic for the operation of a perfect state transfer system in the presence of a systematic error at time $t$ via an initial encoding and corresponding decoding in small regions at either end of the chain.}
\label{fig:q1}
\end{figure}

In order to counteract these systematic errors, we shall encode the information about the state $\ket{\chi}$ in the first $D$ qubits at one end of the chain, which we call the {\em encoding region}. The encoding will be defined so that perfect state transfer can be achieved by applying a unitary decoding operator $U_D$ at time $t_f$, acting just over a {\em decoding region} of size $D$ at the opposite end of the chain (see Fig.\ \ref{fig:q1}). We refer to the region outside the decoding region as the complement region, which has size $\bar D\doteq N-D$. The size $D$ will be determined by the number of errors that we have to encode against, but should be considered small compared to the transfer distance, $N$.

\subsection{State Encoding}

 We propose to encode the information in the state
\begin{equation}
\ket{\psi_{I}}\doteq (\alpha Q_0^{\dg} +\beta Q_1^{\dg})\ket{\psi_0} \label{enc}
\end{equation}
where the encoding operators
\begin{equation}
Q_{a}^{\dagger}\doteq \sum_{i=1}^{D}(\epsilon_i^{a}a_i^{\dagger}+\eta_i^{a}a_i), \;a \in \{0,\;1\}\;,\;\;
\epsilon_i^{a}\;,\;\;\eta_{i}^{a} \in \mathbb{C}\label{eqn:defQ}
\end{equation}
are to be determined to ensure that the decoding operation $U_{D}$ exists; the state $\ket{\psi_0}$ fulfils: 
\begin{equation}
Q_{a}\ket{\psi_0}=0\;, \;\;\forall\;a \in \{0,\;1\}\label{hw1}\;.
\end{equation}
 We require the $\{ Q_{a}\}$'s to represent two orthogonal fermionic modes, which imposes the conditions
\begin{equation}
\left\{  Q_{a}, { Q_{a ^{'}}}^{\dagger}\right\}= \delta_{a, a^{'}} \iff  \sum_{i=1}^{D} (\epsilon_{i}^{a}{\epsilon_{i}^{a^{'}}}^{*}+ \eta_{i}^{a}{\eta_{i}^{a^{'}}}^{*}) = \delta_{a, a^{'}}\;,\;\forall\;a \in \{0,\;1\} \label{antic1}
\end{equation}

\begin{equation}
\left\{  Q_{a}^{\dagger}, { Q_{a ^{'}}}\right\}^{\dagger}= 0 \iff  \sum_{i=1}^{D} (\epsilon_{i}^{a}{\eta_{i}^{a^{'}}}+ \eta_{i}^{a}{\epsilon_{i}^{a^{'}}}) = 0\; .\;\forall\;a \in \{0,\;1\}\label{antic2}
\end{equation}
This ensures that the logical 0, $Q^\dagger_0\ket{\psi_0}$, and the logical 1, $Q^\dagger_1\ket{\psi_0}$, are orthogonal, normalised states, $\bra{\psi_0}Q_{a}{Q_{{a}^{'}}^{\dagger}\ket{\psi_0}}=\delta_{a,a^{'}}$. The idea behind the proposed definition is that if the encoding operators correspond to fermionic modes different from the ones affected by the error, the orthogonality of the encoding states would be unaffected by the error, and so would the encoded information. In addition, since the Hamiltonian describes independent fermions, the transfer of the information-carrying excitations would be independent of the excitations introduced by the error, making it possible to recover the information by applying an appropriate decoding operation at time $t_f$. We shall now formalise this intuition. 

\subsection{State Decoding}

 Let us rewrite  $\ket{\psi_{F}}$, highlighting the action of the time-evolved error and encoding operators on the decoding region.  To do this, we describe the action of any given fermionic operator as a product of the part that acts on the decoding region, and the part that acts on its complement.
\begin{eqnarray*} 
a^{\dagger}_j&=&\tilde f_j^{\dg} \otimes \identity^{\otimes{D}}\;,\;\tilde f_j=\frac{1}{2} Z^{\otimes{j-1}}\otimes\big (X-iY\big) \otimes \identity^{\otimes \bar D-j}\;, \;\forall j\leq \bar D \\ 
 a_j^{\dagger}&=&{ Z}^{\otimes \bar {D}} \otimes f_j^{\dagger}\;, f_j^{\dagger}\doteq \frac{1}{2}Z^{j-\bar D-1}\otimes\;\big (X-iY\big)\otimes \identity^{N-j} \;\;\forall j> \bar D \;,
 \end{eqnarray*}
with $\{f_{j}^{\dagger}, f_i\}=\delta_{i,j}$, $\{f_{i}, f_j\}=0$ (and similarly for the $\tilde f_j$'s).  %Since $f_i(t)=\tilde E_i(t)+E_i(t)$, $\{E_i,E_j\}=0=\{\tilde E_i,\tilde E_j\}$, while $\{E_i,E_j^{\dg}\}= \delta_{i,j}-\{\tilde E_i,\tilde E_j^{\dg}\}$.
Hence, a fermionic operator evolved in time from when it acts up to the decoding time, $t_f$, is expressed as
$$\hat a_n(t)= \tilde F_{n}(t)\otimes \identity^{\otimes{D}}+Z^{\otimes \bar D} \otimes F_{n}(t)\;,$$ with $\displaystyle{\tilde F_{n}(t)\doteq \sum_{m=1}^{\bar D}\beta_{n,m}
(t-t_f) \tilde f_{m}}$, $\displaystyle{F_{n}(t)\doteq \sum_{m=\bar D +1}^{N}\beta_{n,m}(t-t_f) f_{m}\;.}$
As we saw in Sec.\ \ref{sec:moder}, the $\tilde F_n(t)$ are essentially irrelevant, and it is the $F_n(t)$, the parts acting on the decoding region, that we must take care of. Substituting in Eq.\ \eqref{EVER}, one obtains
\begin{equation}
{\cal E}(t_f-t)=\sum_{i=1}^{s}  \gamma_i \sum_{x\in\{0,1\}^{n_i}}\tilde P_{\bar x}^{(i)}\Lambda_{ x}\otimes P_{x}^{(i)} \;\;  \label{REDER}
\end{equation}
summing over the string $x$ in order to convey within the $P_x^{(i)}$ all the different combinations of $F_{k_j^{i}}$ acting on the decoding region 
$$ \;\; P_x^{(i)}\doteq \prod_{j=1}^{m_i}\left(   F_{k_j^{i}}^{\dg}\right)^{x_j} 
\prod_{j=m_i+1}^{n_i}\left( F_{k_j^{i}}\right)^{x_j}\;,\;\;\Lambda_x \doteq (-1)^{\epsilon_{\bar x}}\prod_{j=1}^{n_j}({Z}^{\otimes\bar D})^{x_j},
$$
with $\tilde P_x^{(i)}$ (obtained by substituting $\tilde F_{k_j^{i}}$ for $F_{k_j^{i}}$) conveying the equivalent information for the complement region\footnote{The argument $t$ has been omitted for simplicity.}. The complement of $x$ is denoted by  $\bar x$. $\epsilon_x$ conveys the parity of the ordering of the string $x$, as arises from imposing a standard ordering of the fermionic operators.
Rewriting the encoding operators in the same formalism, one has:
\begin{equation}
U(t_F)\ket{\psi_{I}}= (\alpha \hat{ Q}_0(t_f)+  \hat{ Q}_1(t_f)) \ket{\psi_{out}} \;, \; \ket{\psi_{out}}\doteq U(t_f)\ket{\psi_{0}}\label{REDENC}
\end{equation}
where $$\hat {Q}_{a}^{\dagger}(t_f)={Z}^{\otimes {\bar D}}\otimes q_{a}\;,\;\;q_{a}^{\dagger}\doteq \sum_{i=1}^{D}\epsilon_i^{a}f_{N-i+1}^{\dagger}+\eta_i^{a}f_{N-i+1}\;.$$ 
To complete the rewriting, we note that it is always possible to write $\ket{\psi_{out}}=\sum_{i}\ket{\phi_i}\otimes\ket{\psi_i}$, where $\ket{\phi_i}$ is any (not normalised) state defined over the complement of the decoding region, and, via Eq.\ \eqref{hw1}, $\ket{\psi_i}$ is a solution to
\begin{equation}
q_{a}\ket{\psi}=0\;,  a= 0,1\;. \label{factor}
\end{equation}
One can find this state as an eigenstate of the operator $q_0^{\dagger}q_0q_1^{\dagger}q_1$ with eigenvalue $1$. Provided $D\geq2$, such eigenstates always exist (the size of the Hilbert space spanned by the solutions of this equation is $2^{D-2}$; hence $\ket{\psi_{out}}$ can be chosen in a Hilbert space of dimension $2^{N-2}$, i.e., condition \eqref{hw1} is not too restrictive). In what follows, we shall work with  $\ket{\psi_{out}}=\ket{\phi}\otimes\ket{\psi}$ for the sake of simplicity, and more general states follow by linearity.

The state at time $t_f$ can be rewritten as:
\begin{equation}
\ket{\psi_{F}}=   \sum_{i=1}^{s}\gamma_i\left (\sum_{x\in\{0,1\}^{n_i}} \ket{\phi_{\bar x}^{i}}\otimes (\alpha 
\ket{{\bf 0}^{(i)}_{ x}}+\beta \ket{{\bf 1}^{(i)}_{x}})\right )\;,\label{final}
\end{equation}
with $\ket{{\bf 0}^{(i)}_{x}}\doteq P^{(i)}_{x}q_0\ket{\psi}\;,\;\;\ket{{\bf 1}^{(i)}_{x}}\doteq P^{(i)}_{x}q_1\ket{\psi}\;
$ and $\ket{\phi_{\bar x}^{i}}\doteq \tilde P^{(i)}_{\bar x}\Lambda_x Z^{\otimes{\bar D}}\ket{\phi}$. 
%
%
% Our plan is now to define the operators $Q_0$ and $Q_1$ so that if the encoded states $\bar Q_0\ket{\psi_0}$ and $\bar Q_1\ket{\psi_0}$ are mutually orthogonal, then $\forall x, y$ and $\forall i,j$ so are $\ket{{\bf 0^{(i)}_{ x}}}\;\;,\;\;\ket{{\bf 1^{(j)}_{y}}}$. We shall then  prove that whenever this property holds and the error $\er$ is trace preserving, there exists a unitary operator which allows one to recover perfectly from the error. 
We shall now prove that if Eqns.\ \eqref{antic1}, \eqref{antic2} and \eqref{factor} hold, it is sufficient to impose that 
\begin{eqnarray}
\left\{ q_{a}^{\dagger}, F_{k^{(i)}_j}\right\}&=&0 \iff  \sum_{l=1}^{D} \beta_{k^{(i)}_j\!,l}(t)\eta_{l}^{a}=0 
\nonumber \\
\left\{ q_{a}^{\dagger}, F_{k^{(i)}_j}^{\dagger}\right\}&=& 0 \iff  \sum_{l=1}^{D} \beta^{*}_{k^{(i)}_j\!,l}(t){\epsilon_{h}^{a}}=0  \;,\;\;
\label{antic3}
\end{eqnarray}
for both encoded states ($a=0,1$), and each of the possible operators $F_j$ ($j=1\ldots n_i$) from all possible error strings ($i=1\ldots s$), to ensure the existence of least one unitary $U_{D}$ that,  applied at time $t_f$, perfectly recovers the encoded information. Fulfilling these conditions then defines the coefficients of the encoding operators in Eq. \eqref{eqn:defQ}, completely specifying our strategy.

 Let us introduce the sets of vectors: $Z_{a}\doteq \left \{ \ket{{\bf a}^{(i)}_x}\;: \;x \in\{0,1\}^{n_i}\;\;\forall\,i=1\ldots s\right \}$, $a=0,1$, representing the domain of $U_{D}$ . 
Conditions \eqref{antic3} imply the crucial property \begin{equation} 
  P^{(i)}_x q_{a}=(-1)^{w_x}q_{a}P^{(i)}_x  \;,\;\;\,\; P^{(i)}_x q_{a}^{\dagger}=(-1)^{w_x}q_{a}^{\dagger}P^{(i)}_x\;\;,\;\;\forall a\;, \;\forall x \;,\;\forall i\;.\label{QP}
\end{equation}
Consequently, via \eqref{antic2} and \eqref{factor}, no matter what the error, the two logically encoded states remain unambiguously distinguishable, $
\braket{{\bf 0}^{(i)}_x}{{\bf 1}^{(j)}_y}=0\;$, $\forall x \in 
\{0,1\}^{n_i}$, $\forall y  \in \{0,1\}^{n_j}\;,\; \forall i,j$. This is already sufficient to imply the existence of $U_{D}$, although we give an explicit construction in the Appendix.

We have just proven that conditions \eqref{antic1}, \eqref{antic2}, \eqref{factor} and \eqref{antic3} define a fermionic encoding that protects perfectly from the error, allowing the perfect recovery of the information. The extension of the above procedure 
to the case of errors acting at different times $\{t_i\}$ is straightforward, by defining the sets $\{F_{\ell}(t_i)\}$ and imposing 
for each $t_i$ conditions \eqref{antic3}. Suppose $\bar n$ is the number of distinct sites affected by the error $\er$. There 
are $\displaystyle{4\bar n+6}$ conditions for the $4D$ parameters defining the $q_{a}$'s. The minimal $D$ for which a 
solution may be found is $D=\bar n+2\;$, with $2\leq D\leq N$. Under this assumption, 
conditions \eqref{antic3}, involving operators representing the error-action just on the decoding region, suffice to define fermionic modes 
globally unaffected by the error. 

  In a practical sense, the procedure is worthwhile only if $\bar n\ll N$, since this 
confines the encoding and decoding procedures to the ends of the chain, preserving the central tenet of the 
state transfer protocol. We therefore take the condition $\bar n\ll N$ as defining the concept of a low error rate in this scenario. It is important to emphasise, however, that this counting is in terms of the fermionic description of the error operators. Such counting is favourable for errors such as $Z_i$ or $X_iX_{i+1}$, but there are other local Pauli errors, such as $X_i$, which are necessarily described in terms of $O(i)$ fermionic modes, and therefore may not be included in the low rate condition.

As previously mentioned, the typical assumption that the chain is initialised in some global initial state $\ket{\psi_0}$ is not necessary \cite{KAY,PAT2,PAT}, and so it is in our procedure. Indeed, there are two contexts in which our results apply equally. Either we can choose what the initial state should be, as we have so far assumed, or we are given a fixed initial state, in which case \eqref{factor} gives two additional constraints on the encoding operators, which can be accommodated by our choice of $D$. The advantage of choosing a particular initial state is that it may reduce the effective number of errors that we have to correct for. The following section provides an illustration of this idea.

\section{Determining the Error} \label{sec:determine}

Given a spin-chain affected by an (unknown) error $\er$ acting at time $t$, conditions \eqref{antic3} indicate what knowledge of the error is needed to apply our encoding procedure.  What experiment can one perform in order to gain this knowledge?  One may, of course, use process tomography \cite{NIECHU} to determine $\er(t_f-t)$, but this is extremely inefficient and would provide a lot of redundant information. Our description of the error suggests indeed that there must be more efficient procedures, based on preparing the chain in a suitable set of states and then applying state tomography just on the decoding region to determine the error modes $\{F_k\}$, with little regard for those acting on the complement region, $\{\tilde F_k\}$.
We shall now define a probing procedure in the case $\ket{\psi_0}=\ket{\bf 0}$, motivating the existence of such procedures by using the most commonly assumed fixed initial state.

In the case where the entire chain is initialised in the state $\ket{\psi_0}=\ket{\bf 0}$, it is only necessary to reconstruct a subset of the error modes. Indeed, observe that the 
encoding operators contain only creation operators, i.e., $\eta_i^{a}=0$, $\forall i=1,\ldots D$, hence  in \eqref{antic3} the 
equations $\{q_{a}, F_{k_j^i}\}=0$, $\forall i,j$ are automatically satisfied. Also, note that in $\mathcal{E}$ the fermionic operators are ordered so that the annihilation operators act first. If $\ket{\psi_0}=\ket{\bf 0}$, there are no excitations on the complement region and therefore all error strings with any annihilation operators $\tilde F_k$ on the complement region just give a zero-contribution to the final state. In other words,\ from Eq.\ \eqref{final}, for any $i$, $\ket{\phi_{\bar x}^{i}}\neq0$ if and only 
if $x \in C$, with $C\doteq \{x: \bar x_j=0\;,\forall \;j\geq m_i+1\}$. These $x$ correspond to $\tilde P^{(i)}_{\bar x}$ including no $\tilde F_k$. Furthermore, $\forall x \in C$, since $q_a^{\dg}\ket{0}^{\otimes D}$ belongs to the $1$-excitation subspace, only the error modes containing no more that one annihilation operator $F_k$ acting on the decoding region do not annihilate the encoding state. Namely, \ $P_x^{(i)}q_a^{\dg}\ket{\psi}\neq 0$ if and only if  $n_i=m_i$ or $n_i=m_i+1$. Let us define the sets $S_a\doteq \{i: n_i-m_i=a\}$. We have just shown that to protect against the error it is sufficient to protect against all the  $P_x^{(i)}$ such that  $i\in S_a$, $a=0,1$, and $x\in C$. So, it is sufficient that the encoding operators satisfy:
\begin{eqnarray*}
\{q_{a}^{\dagger}, F_{k_j^i}\}&=&0, \forall j=1\ldots n_i, \forall i\in S_0 \\
\{q_{a}^{\dagger},F_{k_{j}^i}\}&=&0, \forall j=1\ldots n_i, \forall i\in S_1.
\end{eqnarray*}
 Moreover, $\forall i\in S_1$, the latter reduces to just
$$
\{q_{a}^{\dagger},F_{k_{n_i}^i}\}=0, \forall i\in S_1
$$
because this implies that $P_x^{(i)} q_{a}^{\dg}\ket{0}^{\otimes D}=0\;,\; \forall x \in C$, so imposing the conditions for $j=1\ldots m_i$ becomes unnecessary.

In order to provide the information relevant to these conditions, we will probe the system with two different initial states, $\ket{\bf 0}=\ket{0}^{\otimes N}$ and $\ket{\bf 1}=\ket{1}^{\otimes D}\ket{0}^{\otimes {\bar D}}$. Note that both of these are prepared using the same fixed initial state outside the encoding region (in which we have the ability to prepare any state), and this fixed state is the same one that will be used for the state transfer protocol. Consider first using $\ket{\bf 0}$. The only error operators that do not annihilate the probing state are the $P_x^{(i)}$ such that $i\in S_0$ and $x \in C$, which indeed do not include any annihilation operators. In order to reconstruct the information about the corresponding ${F_{k^{(i)}}}^{\dagger}$, we can therefore post-select on the decoding region being the the one-excitation subspace, and perform tomography to determine the corresponding density matrix. The span of states is described by the set $V_0\doteq\{F_{k_i}^{\dg}\ket{0}^{\otimes D}\}$, $\forall j=1\ldots m_i, \forall i\in S_0$. Hence, state tomography \cite{NIECHU} applied just to the decoding region allows one to reconstruct ${\rm Span}\{V_0\}$, which is sufficient information to impose $\{q_{a}^{\dagger}, F_{k_j^i}\}=0$, $\forall j=1\ldots n_i, \forall i\in S_0$.

When we use the initial state $\ket{\bf 1}$, the only non-zero contribution to the final state is given by all the $P_x^{(i)}$ $\forall x \in C$, $i=1\ldots s$. Hence, the projection of this state on the $D-1$ excitation subspace of the decoding region at time $t_f$ includes the action of the $P_x^{(i)}$ with $x\in C$ and $ i\in S_{w_x+1}$, i.e., the $P_x^{(i)}$ where the one more $F_k$ operator acts than $F_j^{\dagger}$ operators. The span of these states hence includes those we are interested in, $V_1\doteq \{F_{k_{m_i+1}}\ket{1}^{\otimes D}\}_{ i\in S_1\;}$, but also includes some others, $V'$. This information is sufficient to impose the remaining conditions $\{q_{a}^{\dagger},F_{k_{n_i}^i}\}=0$, $\forall i\in S_1$, as desired, together with additional (unnecessary) conditions, which would protect from the error-components contributing to $V^{'}$. This redundancy is well tolerated, however, since we assume that the number $\bar n$ of the single error-modes is small compared to $N$. 

Overall, this procedure gives a method to determine the relevant errors using only operations on the encoding/decoding regions, and requiring a number of measurements that scales as $O(2^D)$, and $D\simeq \bar n\ll N$, thereby achieving a significant efficiency saving compared to standard process tomography on the whole chain.

\section{Conclusions}

We have presented a protocol, incorporating encoding and decoding procedures, to achieve perfect state transfer in the presence of low rates of 
systematic errors, whose repeatability allows one to learn about their structure. This procedure may be thought of as an error correction optimally tuned to the error. Indeed, it ensures perfect recovery and, in addition, it has the appealing feature that if the number of sites affected by the error is small compared to the dimension of the whole chain, the encoding and decoding 
operators involve just a small number of qubits, preserving the central feature of the ideal state transfer protocol.

The low rate limit may be relaxed slightly, by extending the procedure to include errors with a support greater than the upper bound $n$, by selecting, via the proposed probing procedure, the first $n$ operators $F_k$ having the highest probability of acting on the decoding region and then encoding against them. This would minimise the probability of an unrecoverable error.

 Our formalism can also be applied in more general scenarios, where the action of the error is not instantaneous, since it is sufficient to describe the effect of any error just at the output time, $t_f$. For instance, consider the case of a perturbed 
Hamiltonian, such as $H=H_0+\delta V$, where $H_0$ is the perfect state transfer Hamiltonian while $\delta V(t)$ is a 
perturbation which acts non-trivially only for a short time interval $\delta$ at time $t$. In the interaction picture, the dynamics is determined by $H^{'}(t)=U(t)\delta V U^{\dagger}(t)$. Hence,
$$
{\mathcal E}=U(t_f-t) \exp(-i\int_{0}^{\delta}H^{'}(\tau){\rm d}\tau )U^{\dagger}(t_f-t) \ket{\psi_{out}}.
$$
For small 
$\delta$, $\exp(-i\int_{0}^{\delta}H^{'}(\tau){\rm d}\tau)$ can be written just like the error \eqref{error}, affecting, 
according to the Lieb-Robinson bound \cite{LIE}, only a small number of sites localised around the region where the perturbation acts. Therefore, the proposed encoding may equally be applied to recover perfect state transfer in the presence of this class of perturbations.

In the future, it will be interesting to see how the error correcting capabilities can be developed further, either by encoding on a single chain or across multiple chains, or by using different network topologies for communication. It will also be important to understand if errors such as local bit-flips can be corrected for.

{\em Acknowledgements:} This work is supported by the National Research Foundation and the Ministry of Education, Singapore. CM is supported by EPSRC and the Istituto Superiore Mario Boella.

\appendix

\section{Construction of the decoding unitary}

 To explicitly construct the decoding unitary, one orthogonalises the vectors belonging to each $Z_{a}$ via the Gram-Schmidt procedure:
\begin{equation}
\ket{{\bf \breve a}_{x,j}}\doteq \ket{{\bf a}^{(j)}_{x}}-\sum_{y,l} \frac{ \braket{{\bf \breve a}_{y,l}}{{\bf a}^{(j)}_{x}} }{\braket{{ \bf \breve a}_{y,l}} {{\bf \breve a}_{y,l}}}\ket{{\bf \breve a}_{y,l}} \label{GS0}
\end{equation}
The number $z$ of such vectors satisfies $z\leq \sum_{j=1}^{s}2^{n_j}\;$ (as some vectors in $Z_{a}$ may be linearly dependent). 
Then, $U_{D}$ can be defined as:

$$
U_{D}= \sum_{w,l}\sum_{a} \left(\ket{{w,l}}\otimes\ket{a}\right) \bra{{\bf \breve a}_{w,l}},
$$
where the sum runs over all the orthogonalised vectors and $\{\ket{{w,h}}\}$ is any set of orthonormal vectors, defined over all the decoding region but qubit $N$. 

 To see that the above is the desired unitary, one rewrites $\ket{\psi_{F}}$ in terms of the orthogonalised vectors and uses $$\braket{ {\bf \breve 0}_{w,l}} {{\bf 0}^{(j)}_{x} }=\braket{ {\bf \breve 1}_{w,l}} {{\bf 1}^{(j)}_{x} }\doteq \mu_{w,l}^{(x,j)}\;\;\forall j, x,i, y\;,$$ which holds because \eqref{antic1}, \eqref{factor} and \eqref{QP} imply $$\braket{{\bf 1}^{(i)}_x}{{\bf 1}^{(j)}_y}=\braket{{\bf 0}^{(i)}_x}{{\bf 0}^{(j)}_y}\;,\;\forall x, \forall y \;,\;\forall i,j\;.$$ Consequently, $U_{D}$ applied to $\ket{\psi_{F}}$, gives:
$$
\ket{\psi_{d}}\doteq
\left ( \identity_{\bar D}\otimes U_{D}\right )\ket{\psi_F}=    \ket{\Phi}\otimes 
\left(\alpha{\ket {{0}}}_{N}+\beta {\ket {{1}}}_{N}\right),
$$
 where  $\ket{\Phi}\doteq \sum_{x\in\{0,1\}^{n_i}} \ket{\phi_{\bar x}^{i}}\otimes \sum_{w,l} \mu_{w,l}^{x,i} \ket{{w,l}}$. Since ${\cal E}$ is trace-preserving,
$$
{\rm Tr}_{1\ldots N-1}(\ket{\psi_{d}}\bra{\psi_d})= \ket{\chi}\bra{\chi},
$$
i.e., the information has been perfectly transferred to the last qubit, as promised.

\end{document}